\documentclass[10pt,twocolumn,secnumarabic,amssymb,nobibnotes,aps,prl]{revtex4-2}
\usepackage[T1]{fontenc}
\usepackage{graphicx}
\usepackage{amsmath}

\begin{document}

\title{Robust q-negative Multifractal Detrended Cross-Correlation Coefficient}%

\author{Thiago B. Murari}
\email{thiago.murari@fieb.org.br}
\affiliation{Universidade SENAI CIMATEC, Computational Modeling, Av. Orlando Gomes 1845 (Piat\~a), Salvador, BA 41650-010, Brazil}

\author{Jos\'e Fernando F. Mendes}
\email{jfmendes@ua.pt}
\affiliation{Universidade de Aveiro, Campus Universit\'ario de Santiago, Aveiro 3810-193, Portugal}

\author{Hernane B. B. Pereira}
\email{hernane@fieb.org.br, hbbpereira@gmail.com}
\affiliation{Universidade SENAI CIMATEC, Computational Modeling, Av. Orlando Gomes 1845 (Piat\~a), Salvador, BA 41650-010, Brazil}
\affiliation{UNEB, DEDC, Av. Silveira Martins 2555, Salvador, BA 41195-001, Brazil}

\author{Marcelo A. Moret}
\email{moret@fieb.org.br, mamoret@gmail.com}
\affiliation{Universidade SENAI CIMATEC, Computational Modeling, Av. Orlando Gomes 1845 (Piat\~a), Salvador, BA 41650-010, Brazil}
\affiliation{UNEB, DCET - Physics, Av. Silveira Martins 2555, Salvador, BA 41195-001, Brazil}

\begin{abstract}

The multifractal detrended cross-correlation coefficient $\rho_q(n)$ is widely used to investigate scale-dependent interactions, but its application to negative fluctuation orders is affected by numerical instabilities, unbounded values, and interpretational difficulties. We propose a Signed Multifractal Detrended Cross-Correlation Coefficient, \(\rho_{\mathrm{SMFDCCA}}(n,q)\), an amplitude-conditioned correlation observable for multifractal detrended analysis, based on locally normalized detrended correlations and regularized fluctuation amplitudes.The proposed coefficient preserves the sign of local interactions, remains strictly bounded within $[-1,1]$ for both positive and negative values of $q$, and eliminates the corrective procedures required by previous approaches. Validation using independent fractional Gaussian noise confirms the absence of spurious cross-correlations and the numerical stability of the method. 
Applications demonstrate that the proposed observable resolves how cross-correlations evolve jointly with temporal scale and fluctuation amplitude, revealing scale- and amplitude-dependent correlation structures, including stronger synchronization during large fluctuations in stock-market indices and heterogeneous coupling patterns in temperature records.

\end{abstract}

\maketitle

\section{Introduction}\label{sec1}

Numerous systems, composed of a large number of interacting elements organized into structures that may exist and coexist across multiple scales, exhibit emergent properties that cannot be observed in simpler systems. These elements often interact through nonlinear mechanisms, giving rise to collective behaviors and patterns that cannot be inferred from the properties of individual components \cite{di2025epj}.

A deeper understanding of the dynamics of complex time series requires the use of analytical tools that go beyond traditional approaches based on stationarity and linearity assumptions. To address these challenges, the Detrended Fluctuation Analysis (DFA) method was developed \cite{peng1994mosaic}. DFA has become a widely used technique for the determination of fractal scaling properties and the detection of long-range autocorrelations in noisy, nonstationary time series.

While DFA focuses on the autocorrelation of a single signal, there are many situations where different variables are recorded simultaneously and exhibit interdependencies. To quantify long-range cross-correlations in the presence of nonstationarity, the Detrended Cross-Correlation Analysis (DCCA) was proposed \cite{podobnik2008detrended}. DCCA is a generalization of DFA based on detrended covariance, designed to investigate power-law cross-correlations between two different time series of equal length.

However, many records require more than a single scaling exponent to be fully described, due to different scaling behaviors of small and large fluctuations. To account for this, the Multifractal Detrended Fluctuation Analysis (MFDFA) was introduced as a generalization of the DFA method. MFDFA enables the reliable multifractal characterization of nonstationary time series without requiring computationally heavy modulus maxima procedures \cite{kantelhardt2002multifractal}.

Later, the DCCA method was also generalized to unveil the multifractal features of two cross-correlated signals. This method, known as Multifractal Detrended Cross-Correlation Analysis (MFDCCA or MF-DXA), was proposed to investigate multifractal behaviors in the power-law cross-correlations between two nonstationary time series or higher-dimensional quantities \cite{zhou2008multifractal}.

Despite the utility of DCCA and MFDCCA in identifying power-law dependencies, these methods alone did not directly quantify the level or strength of the cross-correlations. To solve this, was proposed the DCCA cross-correlation coefficient ($\rho_{DCCA}$) \cite{zebende2011dcca}. The $\rho_{DCCA}$ coefficient is defined as the ratio between the detrended covariance function and the detrended variance functions. It provides a dimensionless measure ranging from -1 to 1, quantifying the level of cross-correlation in nonstationary time series, such as financial data \cite{nascimento2018cross,ferreira2020dcca}, mobility \cite{azevedo2021sustainable} and epidemics \cite{oliveira2023paradox,figueredo2023analysis}.

Building upon $\rho_{DCCA}$ method, it was proposed the multifractal cross-correlation coefficient ($\rho_q(n)$) \cite{kwapien2015detrended}. This coefficient generalizes $\rho_{DCCA}$ into a multifractal framework to quantify the degree of cross-correlation between two nonstationary time series across different scales as a function of the fluctuation order $q$. In the $\rho_q(n)$ method, the parameter $q$ acts as a filter: for $q > 2$, it gives more weight to large fluctuations, whereas $q < 2$ emphasizes small fluctuations, allowing researchers to map whether cross-correlations are homogeneous across all amplitudes or dominated by specific fluctuation magnitudes.

Recently, was proposed the Multifractal detrended multiple cross-correlation coefficient ($qDMC^2_x(n)$) \cite{silvafilho2026multifractal}. This coefficient constructs a methodology by integrating the multifractal cross-correlation coefficient $\rho_q(n)$ with the standard multiple correlation coefficient $DMC^2_x(n)$.

But a critical issue associated with multifractal detrended cross-correlation methods emerges for negative values of the fluctuation order parameter \(q<0\). In multifractal analysis, negative \(q\)-orders amplify the contribution of segments characterized by very small fluctuation amplitudes, while positive \(q\)-orders emphasize large fluctuations. Although this property is theoretically useful for investigating weak fluctuations, it introduces important mathematical and numerical instabilities.

According to the discussion presented by Kwapień et al.~\cite{kwapien2015detrended}, the main difficulty associated with negative values of the fluctuation order parameter (\(q<0\)) arises from the excessive emphasis placed on segments containing very small fluctuations. While positive values of \(q\) highlight large-amplitude fluctuations, negative values strongly amplify low-amplitude regions, making the coefficient extremely sensitive to weak signals, noise, and numerical inaccuracies. As a consequence, the multifractal cross-correlation coefficient may lose its conventional interpretation as a bounded correlation measure, since values larger than unity in magnitude can emerge for \(q<0\). Kwapień et al. explicitly describe the interpretation of negative-\(q\) results as a delicate issue, emphasizing that the coefficient may exhibit unstable and erratic behavior in this regime. To reduce this problem, they introduced an additional inversion procedure whenever the coefficient exceeds the expected bounds. Therefore, the original qDCCA formulation does not intrinsically resolve the instability associated with negative fluctuation orders, but rather compensates for it through a corrective post-processing step.

The same limitation propagates directly into the recently proposed multifractal detrended multiple cross-correlation coefficient \(qDMC_x^2(n)\) developed by Silva Filho et al.~\cite{silvafilho2026multifractal}, since the method is constructed explicitly from the original \(\rho_q(n)\) coefficient. Consequently, the numerical instability and boundedness problems associated with \(q<0\) remain embedded in the multivariate framework. It is important to note that Sierra-Porta~\cite{sierraporta2025multifractal}, in the analysis of cosmic ray intensity and sunspot number using the multifractal detrended cross-correlation coefficient \(\rho_q(n)\), restricted the investigation to positive fluctuation orders (\(q \geq 0\)). By excluding negative \(q\)-values, Sierra-Porta avoided these instability issues and restricted the analysis to fluctuation regimes where the coefficient remains more numerically robust and physically interpretable.


Despite the advances provided by the multifractal detrended cross-correlation coefficient \(\rho_q(n)\) and its multivariate extension \(qDMC_x^2(n)\), some mathematical and numerical limitations remain unresolved, particularly for negative fluctuation orders (\(q<0\)). Such difficulties become relevant when studying systems characterized by weak fluctuations, intermittency, or near-constant local segments. Motivated by these limitations, the present work introduces a amplitude-conditioned correlation observable for multifractal detrended analysis, namely the Signed Multifractal Detrended Cross-Correlation Coefficient, based on locally normalized signed correlations and regularized fluctuation amplitudes. In the present context, the term multifractal refers to the q-dependent weighting of fluctuation amplitudes, which enables the characterization of cross-correlations associated with different fluctuation magnitudes. The proposed observable provides a bounded measure of amplitude-conditioned cross-correlations across temporal scales for both positive and negative values of \(q\).

\section{Signed Multifractal Detrended Cross-Correlation Coefficient}\label{sec2}

Let \(x_i\) and \(y_i\), with \(i=1,\ldots,N\), be two time series of equal length. The integrated profiles are first constructed as

\begin{equation}
\small
X_k
=
\sum_{i=1}^{k}
\left(x_i-\bar{x}\right),
\quad
Y_k
=
\sum_{i=1}^{k}
\left(y_i-\bar{y}\right),
\quad
k=1,\ldots,N.
\end{equation}

The profiles are divided into \(N_n\) segments of equal length \(n\). In each segment \(v\), polynomial trends of order \(m\) are fitted to the profiles:

\begin{equation}
\widetilde{X}_{v,m}(k),
\qquad
\widetilde{Y}_{v,m}(k),
\qquad
k=1,\ldots,n.
\end{equation}

The detrended residuals are then obtained as

\begin{equation}
\varepsilon_{x,v}(k)
=
X_v(k)-\widetilde{X}_{v,m}(k),
\end{equation}

\begin{equation}
\varepsilon_{y,v}(k)
=
Y_v(k)-\widetilde{Y}_{v,m}(k).
\end{equation}

The local detrended variances are defined by

\begin{equation}
F_x^2(n,v)
=
\frac{1}{n}
\sum_{k=1}^{n}
\varepsilon_{x,v}^2(k),
\end{equation}

\begin{equation}
F_y^2(n,v)
=
\frac{1}{n}
\sum_{k=1}^{n}
\varepsilon_{y,v}^2(k).
\end{equation}

The local detrended covariance is

\begin{equation}
C_{xy}(n,v)
=
\frac{1}{n}
\sum_{k=1}^{n}
\varepsilon_{x,v}(k)\,
\varepsilon_{y,v}(k).
\end{equation}

Unlike the local variances, \(C_{xy}(n,v)\) may assume positive or negative values depending on the direction of the local cross-correlation.

To avoid numerical instabilities and division by zero, particularly for negative values of \(q\), a small positive regularization constant \(\epsilon>0\) is introduced. This parameter acts as a numerical floor that prevents singularities and undefined operations when the detrended variances become extremely small or vanish within a local segment. The local fluctuation amplitude is then defined as

\begin{equation}
A_v(n)
=
\sqrt{
\max\!\left(F_x^2(n,v),\epsilon\right)
\max\!\left(F_y^2(n,v),\epsilon\right)
}.
\end{equation}

The regularization is also essential for the evaluation of logarithmic and power-law weighting terms, especially for negative multifractal orders $q<0$, where small fluctuation amplitudes receive disproportionately large weights. Without regularization, segments with vanishing amplitudes would lead to weights and numerical instabilities. The parameter $\epsilon$ is therefore chosen as a very small constant without loss of generality ($\epsilon=10^{-12}$), so that it remains negligible compared to the characteristic fluctuation amplitudes of the data while still guaranteeing mathematically well-defined and computationally stable calculations. Although $\epsilon$ slightly regularizes the behavior of extremely low-amplitude segments, its influence on the overall multifractal structure is minimal for practical ranges of $q$.

Using this quantity, the local detrended correlation coefficient becomes

\begin{equation}
r_v(n)
=
\frac{
C_{xy}(n,v)
}{
A_v(n)
}.
\end{equation}

By the Cauchy--Schwarz inequality,

\begin{equation}
-1 \le r_v(n)\le 1.
\end{equation}

For a multifractal order \(q\in\mathbb{R}\), the weight assigned to each segment is

\begin{equation}
w_v(q,n)
=
A_v(n)^{q/2},
\qquad q\neq0.
\end{equation}

For \(q=0\), the limiting case corresponds to equal weighting of all segments:

\begin{equation}
w_v(0,n)=1.
\end{equation}

The signed multifractal detrended cross-correlation coefficient is then defined as

\begin{equation}
\label{eqmain}
\rho_{\mathrm{SMFDCCA}}(n,q)
=
\frac{
\sum_{v=1}^{N_n}
w_v(q,n)\,
r_v(n)
}{
\sum_{v=1}^{N_n}
w_v(q,n)
},
\qquad
q\in\mathbb{R}.
\end{equation}

Equation (\ref{eqmain}) defines an amplitude-conditioned correlation observable obtained as the expectation of bounded local detrended correlation coefficients under a q-dependent weighting of fluctuation amplitudes. Consequently, each value of q probes a distinct fluctuation-amplitude regime while preserving the interpretation of the observable as a bounded signed correlation coefficient. Equivalently,

\begin{equation}
\rho_{\mathrm{SMFDCCA}}(n,q)
=
\frac{
\sum_{v=1}^{N_n}
A_v(n)^{q/2}
\,
\dfrac{
C_{xy}(n,v)
}{
A_v(n)
}
}{
\sum_{v=1}^{N_n}
A_v(n)^{q/2}
},
\qquad
q\neq0.
\end{equation}

It should be emphasized that $\rho_{\mathrm{SMFDCCA}}(n,2)$ does not, in general, reduce algebraically to the conventional DCCA coefficient, since the former is constructed as an amplitude-weighted average of local detrended correlations, whereas the latter is defined from the aggregated detrended covariance and variances.

For \(q=0\), the coefficient reduces to the arithmetic mean of the local detrended correlations:

\begin{equation}
\rho_{\mathrm{SMFDCCA}}(n,0)
=
\frac{1}{N_n}
\sum_{v=1}^{N_n}
r_v(n).
\end{equation}

Since the weights satisfy

\begin{equation}
w_v(q,n)\ge0,
\end{equation}

the coefficient is a weighted average of local correlations and therefore satisfies

\begin{equation}
-1
\le
\rho_{\mathrm{MFDCCA}}(n,q)
\le
1.
\end{equation}

This formulation is valid for positive, negative, and zero values of \(q\). Positive values of \(q\) emphasize segments with larger fluctuation amplitudes, while negative values emphasize smaller fluctuations. Consequently, \(\rho_{\mathrm{SMFDCCA}}(n,q)\) can be interpreted as the expected local detrended correlation conditioned by fluctuation amplitude, providing a physically interpretable observable across different fluctuation regimes. Because the sign of the local covariance \(C_{xy}(n,v)\) is preserved through \(r_v(n)\), the coefficient simultaneously quantifies both the strength and direction of multifractal cross-correlations. 

The \(\rho_{\mathrm{SMFDCCA}}(n,q)\) coefficient is rooted in the same q-dependent fluctuation formalism that underlies MFCCA and $\rho_q(n)$. As discussed by Wątorek \textit{et al.} \cite{watorek2021}, q-dependent correlation coefficients and multifractal cross-correlation exponents provide complementary information about the interaction structure of complex systems, with the parameter $q$ acting as a selective filter for fluctuation amplitudes. By preserving this fluctuation-selective weighting mechanism, \(\rho_{\mathrm{SMFDCCA}}(n,q)\) retains the ability to distinguish correlation structures associated with different fluctuation magnitudes.

It is important to emphasize that the multifractal character of \(\rho_{\mathrm{SMFDCCA}}(n,q)\)arises from its explicit dependence on the fluctuation order q, which defines a family of amplitude-conditioned observables. Consequently, the coefficient provides an amplitude-stratified characterization of cross-correlations. Unlike multifractal scaling frameworks such as MFDFA or MFCCA, the proposed coefficient is not designed to estimate generalized Hurst exponents, multifractal spectra, or singularity distributions. Its purpose is to quantify how the strength and direction of cross-correlations vary across fluctuation magnitudes and temporal scales while preserving boundedness and numerical stability.


\section{Discussion}\label{sec3}

\subsection{Computer Generated Random Signal}

To establish a null benchmark for the proposed observable, we generated an ensemble of one thousand statistically independent pairs of fractional Gaussian noise (fGn) time series using the Davies--Harte algorithm. Each series contains $N=20,000$ data points and has Hurst exponent $H=0.5$, corresponding to an uncorrelated Gaussian process without long-range temporal correlations. No preprocessing or transformation was applied to the generated signals. The two members of each pair were generated independently.

For each of the one thousand independent pairs, $\rho_{\mathrm{SMFDCCA}}(n,q)$ was evaluated using linear detrending ($m=1$), overlapping segmentation, reverse segmentation, and the regularization parameter $\varepsilon=10^{-12}$. The fluctuation orders, integer numbers, from $q=-5$ to $q=5$ were considered. For every $(n,q)$ combination, the ensemble mean, standard deviation, 95\% confidence interval of the mean, and empirical 2.5th and 97.5th percentiles were calculated from the one thousand independent realizations.

Table~\ref{tab:null_benchmark} summarizes the principal null-model results for each fluctuation order. For all investigated values of $q$, the ensemble mean remains statistically centered at zero, with no systematic displacement as a function of either $q$ or the temporal scale. The negative and positive orders therefore produce symmetric null behavior despite emphasizing different fluctuation-amplitude regimes.

\begin{table*}[!htb]
\centering
\caption{Null-model statistics from 1,000 independent fGn pairs with $H=0.5$. The maximum absolute ensemble mean, $\max|\langle\rho\rangle|$, is evaluated over the available scales ($13\leq n\leq4181$). The standard deviations and empirical 95\% null intervals are given for the smallest and largest scales.}
\label{tab:null_benchmark}
\begin{ruledtabular}
\begin{tabular}{cccccc}
$q$ & $\max|\langle\rho\rangle|$ & $\sigma_{n=13}$ & $\sigma_{n=4181}$ & 95\% CI at $n=13$ & 95\% CI at $n=4181$ \\
\colrule
$-5$ & 6.2e-03 & 0.015 & 0.281 & $[-0.030, 0.030]$ & $[-0.551, 0.548]$ \\
$-4$ & 4.9e-03 & 0.013 & 0.245 & $[-0.026, 0.026]$ & $[-0.478, 0.480]$ \\
$-3$ & 6.0e-03 & 0.012 & 0.210 & $[-0.023, 0.024]$ & $[-0.407, 0.413]$ \\
$-2$ & 5.1e-03 & 0.011 & 0.180 & $[-0.022, 0.021]$ & $[-0.354, 0.353]$ \\
$-1$ & 5.4e-03 & 0.010 & 0.160 & $[-0.020, 0.020]$ & $[-0.312, 0.312]$ \\
$0$  & 7.8e-03 & 0.009 & 0.153 & $[-0.018, 0.018]$ & $[-0.300, 0.288]$ \\
$1$  & 8.2e-03 & 0.009 & 0.166 & $[-0.017, 0.017]$ & $[-0.326, 0.307]$ \\
$2$  & 8.6e-03 & 0.011 & 0.183 & $[-0.022, 0.022]$ & $[-0.360, 0.342]$ \\
$3$  & 8.9e-03 & 0.013 & 0.203 & $[-0.025, 0.025]$ & $[-0.401, 0.387]$ \\
$4$  & 1.0e-02 & 0.016 & 0.225 & $[-0.031, 0.031]$ & $[-0.442, 0.426]$ \\
$5$  & 1.2e-02 & 0.019 & 0.247 & $[-0.037, 0.037]$ & $[-0.491, 0.473]$ \\
\end{tabular}
\end{ruledtabular}
\end{table*}

The ensemble of the generated random signal results demonstrates three properties of the proposed coefficient. First, across all investigated scales and fluctuation orders, the ensemble mean oscillates around zero. No systematic bias is observed for either negative or positive $q$. This behavior is consistent with the construction of $\rho_{\mathrm{SMFDCCA}}(n,q)$ as a weighted average of local signed correlations.

Second, the null benchmark demonstrates the stability of the coefficient for negative fluctuation orders. Negative $q$ values assign greater weight to segments with small fluctuation amplitudes and therefore provide a particularly stringent test of the regularized formulation. Nevertheless, the behavior for $q=-5$ remains centered at zero and is qualitatively indistinguishable from that obtained for the corresponding positive order $q=5$. No divergence or systematic displacement is observed as $q$ becomes increasingly negative. This result confirms that the $\varepsilon$-regularization prevents numerical singularities while preserving the expected null behavior.

Third, the coefficient satisfies its theoretical bounds throughout the complete ensemble. Across the one thousand independent pairs, all investigated scales, and all ten fluctuation orders, no value violating $|\rho_{\mathrm{SMFDCCA}}|>1$ was observed.

\subsection{U.S. equity markets}

\begin{figure}[!htb]
\centering
\includegraphics[width=\columnwidth]{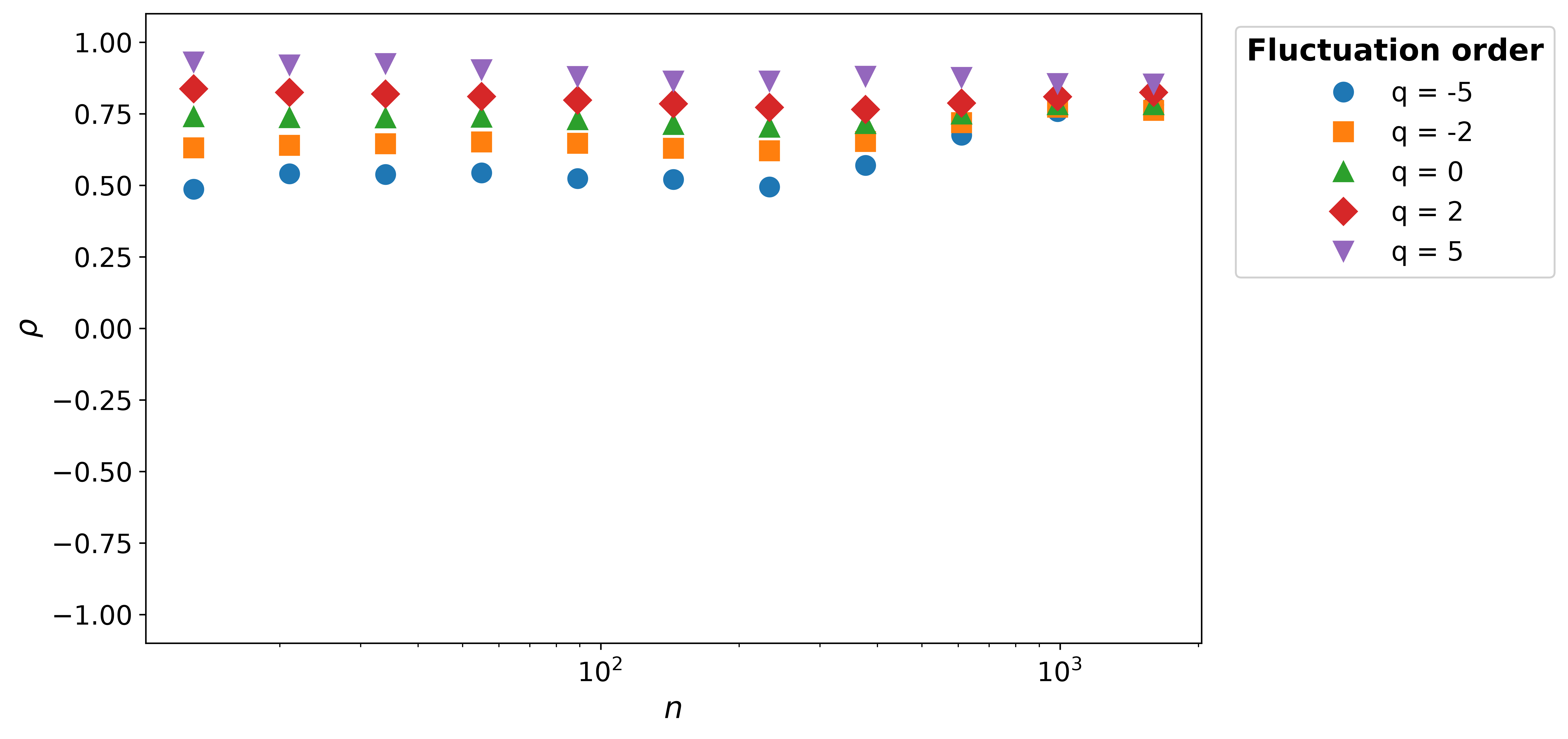}
\caption{\(\rho_{\mathrm{SMFDCCA}}(n,q)\) between the daily logarithmic returns of the Dow Jones Industrial Average and NASDAQ Composite Index from January 2, 1992 to May 8, 2026 ($N=8649$). The coefficient is shown for fluctuation orders $q \in [-5,5]$ as a function of the window size $n$. Positive values of \(\rho_{\mathrm{SMFDCCA}}(n,q)\) across all scales indicate persistent cross-correlations between the two U.S. equity markets. The increase of \(\rho_{\mathrm{SMFDCCA}}(n,q)\) with $q$ reveals strong synchronized co-movements for large-amplitude fluctuations than for small fluctuations, evidencing q-dependent asymmetry in the correlation structure.}
\label{fig2}
\end{figure}

To empirically validate the proposed observable, we evaluate the amplitude-conditioned correlation \(\rho_{\mathrm{SMFDCCA}}(n,q)\) coefficient on the daily logarithmic returns of the Dow Jones Industrial Average (\^{}DJI) and NASDAQ Composite (\^{}IXIC) indices from January 2, 1992, to May 8, 2026 ($N=8649$), collected from Yahoo Finance. The presence of power-law cross-correlations between these two benchmark U.S. markets is a well-established stylized fact in the complex systems literature \cite{podobnik2008detrended, zebende2011dcca}. As illustrated in Figure~\ref{fig2}, our analysis reveals persistent positive cross-correlations across all investigated temporal scales $n$ and fluctuation orders $q \in [-5,5]$, with the coefficient bounded between $0.49$ and $0.93$. We observe a pronounced asymmetry in the correlation structure: positive moments ($q>0$) yield significantly stronger cross-correlations than negative moments ($q<0$). This discrepancy explicitly confirms that large-amplitude fluctuations are characterized by a much higher degree of synchronization than small-amplitude fluctuations, a phenomenon consistent with the non-uniform coupling typically observed in empirical multifractal systems \cite{kwapien2015detrended,Jiang2019}.

A scale dependence is also observed. At short windows, the cross-correlation spectrum exhibits the largest spread, increasing from $\rho_{\mathrm{SMFDCCA}}(13,-5)=0.486$ to $\rho_{\mathrm{SMFDCCA}}(13,5)=0.931$. It may indicate a significant heterogeneity in the q-dependent coupling mechanism, where large return fluctuations are almost synchronized while small fluctuations remain only moderately synchronized \cite{kwapien2015detrended, silvafilho2026multifractal}. As the scale increases, the dependence on $q$ progressively weakens. For the largest investigated window ($n=1597$), the coefficient varies only from $0.782$ to $0.854$, revealing a substantial reduction of the amplitude-dependent asymmetry in the correlation structure. This convergence suggests that long-term market dynamics are dominated by common macroscopic factors that homogenize the correlation structure across fluctuation amplitudes. It is consistent with the presence of nonlinear interactions and scale-dependent coupling mechanisms, which are generally found on complex financial systems \cite{DiMatteo2007, barunik2012understanding, Jiang2019}.

The persistence of strong positive $\rho_q$ values demonstrates multiscale co-movements between the industrial and technology sectors. By deploying the amplitude-conditioned correlation observable \(\rho_{\mathrm{SMFDCCA}}(n,q)\), we can quantitatively present how this interaction strength evolves simultaneously with both temporal scale and fluctuation magnitude. These findings underscore the nonlinear, hierarchical organization of modern financial markets \cite{Mantegna1999, Jiang2019}, proving the efficacy of the proposed observable in characterize multiscale co-movements. \(\rho_{\mathrm{SMFDCCA}}(n,q)\) may also provide an amplitude-resolved view of the scale-dependent synchronization traditionally associated with the Epps effect \cite{epps1979comovements,drozdz2010foreign}.

\subsection{Weather data}

The meteorological data were obtained from the Banco de Dados Meteorológicos para Ensino e Pesquisa (BDMEP) maintained by the Brazilian National Institute of Meteorology (INMET). Daily observations were collected from the São Paulo, Mirante weather station (Station A701; latitude $23.4963^{\circ}$ S, longitude $46.6201^{\circ}$ W; altitude 785.64~m). The analyzed dataset covers the period from July 24, 2006, to May 31, 2026, and includes daily measurements of mean air temperature, mean relative humidity, maximum temperature, and minimum temperature ~\cite{INMET_BDMEP}.

\begin{figure}[!htb]
\centering
\includegraphics[width=\columnwidth]{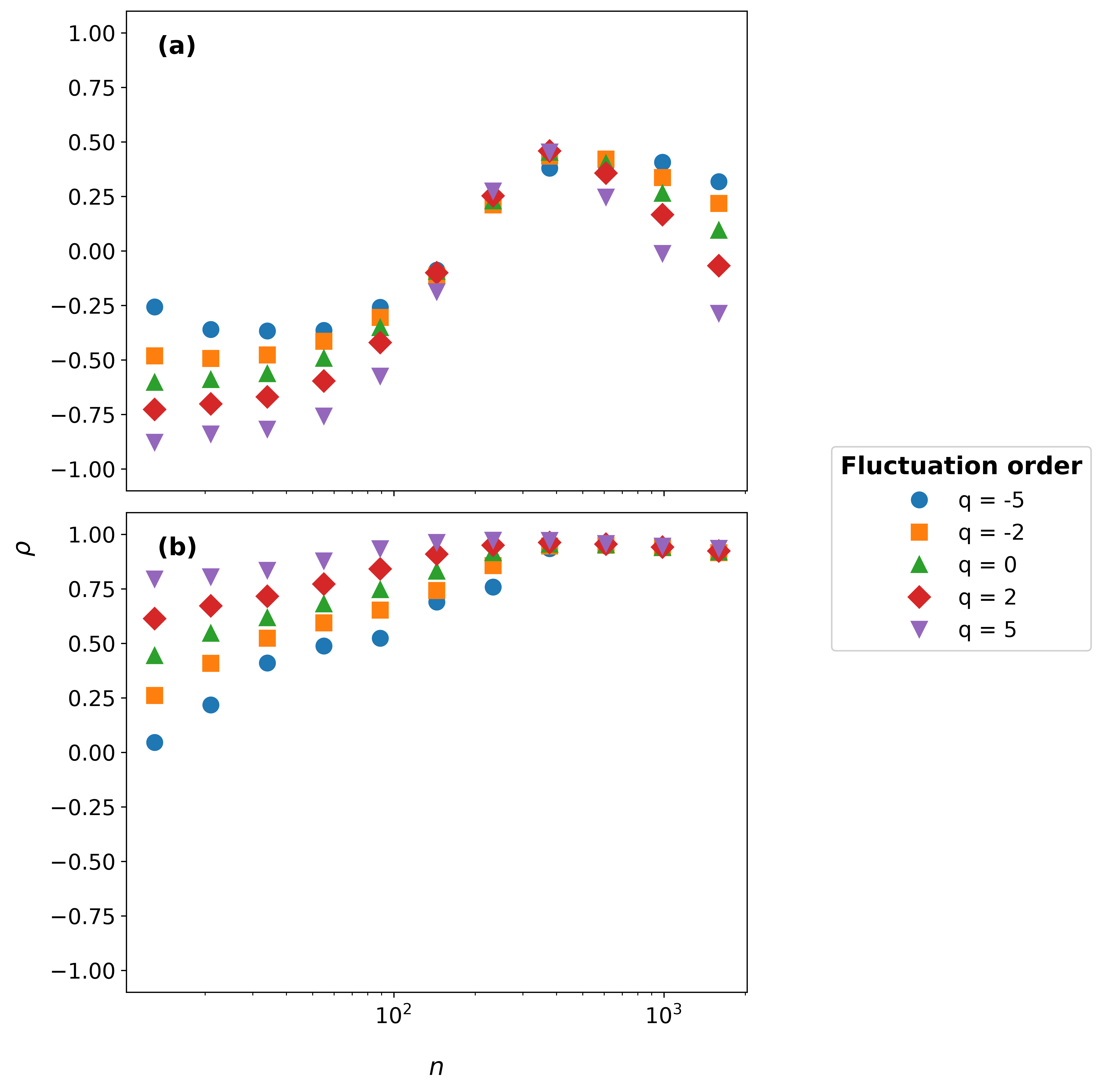}
\caption{
\(\rho_{\mathrm{SMFDCCA}}(n,q)\), for meteorological variables in São Paulo, Brazil. (a) Mean temperature versus mean relative humidity in São Paulo. (b) Maximum versus minimum temperature in São Paulo. The dependence on the fluctuation order $q \in [-5,5]$ reveals heterogeneous cross-correlation structures across scales.}
\label{fig3}
\end{figure}

The weather graphs displayed in Figure~\ref{fig3} reveal different cross-correlation structures depending on the climatic variable pair. For the mean temperature--mean humidity relationship, São Paulo (Fig.~\ref{fig3}a) exhibit strong negative cross-correlations at short temporal scales, indicating that local increases in temperature are generally associated with decreases in relative humidity. The coefficient evolves from strongly negative values at short scales toward positive values at intermediate scales before partially reverting at the largest windows, suggesting a complex dynamic between local atmospheric processes and seasonal forcing.

The maximum temperature--minimum temperature pairs (Fig.~\ref{fig3}b) exhibit a different co-movement behavior. \(\rho_{\mathrm{SMFDCCA}}(n,q)\) remains positive across almost all scales and fluctuation orders, indicating persistent coupling between daily temperature measurements. A strong comovement observed in São Paulo, where the coefficient approaches unity at intermediate and large scales, indicating a very strong coupling between maximum and minimum temperatures. Positive fluctuation orders produce larger coefficients than negative orders, indicating that large temperature fluctuations are more strongly synchronized than weak fluctuations.

\subsection{Limitations}
Although the proposed observable is mathematically bounded and preserves the sign of local cross-correlations, extreme positive and negative values of q may induce numerical and statistical instabilities. The overall cross-correlation coefficient is derived from a weighted average of local cross-correlations $r_i$. The weight $W_i$ of each local segment $i$ is proportional to its fluctuation amplitude raised to the power of $q$:
\begin{equation}
    W_i \propto (\text{Amplitude}_i)^q
\end{equation}

This is the reason the method is sensitive to localized variance spikes when evaluated at these high moment orders, particularly at large temporal scales where the total number of segments is inherently small. Consequently, practical applications may employ moderate q-ranges.

For daily or low-frequency financial return series, empirical guidelines suggest restricting the range of $q$ to $|q| \le 6$, or even smaller, to preserve estimation accuracy \cite{Jiang2019}. When dealing with heavy-tailed datasets such as high-frequency equity or cryptocurrency returns (where the tail index $\gamma \approx 3$), the range of $q$ is typically restricted to $[-4, 4]$ due to the non-existence of higher-order moments in the underlying probability density functions \cite{watorek2021}. Furthermore, for certain polynomial or parametric fits of scaling functions, researchers often restrict the analysis to $q \ge -1$ because when the empirical tail index satisfies $\gamma > 2$, moments of returns for $q < -1$ do not formally exist \cite{Jiang2019}.

\section{Summary}\label{sec4}

This work introduces the Signed Multifractal Detrended Cross-Correlation Coefficient, \(\rho_{\mathrm{SMFDCCA}}(n,q)\) as an amplitude-conditioned correlation observable within the multifractal detrended analysis framework, for quantifying multifractal cross-correlations while preserving both the strength and direction of local interactions. Unlike previous multifractal cross-correlation coefficients, the proposed formulation is based on locally normalized detrended correlations and positive-definite fluctuation weights, resulting in a bounded coefficient within the interval $[-1,1]$ for all values of the fluctuation order $q$. The method addresses a well-documented limitation of conventional q-dependent detrended cross-correlation approaches, namely the numerical instability and loss of interpretability associated with negative fluctuation orders. By introducing a regularized local fluctuation amplitude and performing the averaging directly on bounded local correlation coefficients, the proposed approach eliminates the need for corrective inversion procedures and provides a unified treatment of positive and negative moments.

The proposed observable was validated through a combination of synthetic and empirical applications. For independent fractional Gaussian noise series generated with $H=0.5$, the coefficient remained close to zero for all scales and fluctuation orders, confirming the absence of spurious cross-correlations and demonstrating numerical robustness in both positive and negative $q$ regimes. Application to the daily logarithmic returns of the Dow Jones Industrial Average and NASDAQ Composite indices revealed persistent positive cross-correlations across all investigated scales, with stronger synchronization for large-amplitude fluctuations than for small fluctuations. These findings demonstrate that the proposed coefficient successfully captures scale-dependent and amplitude-dependent coupling structures while preserving a straightforward interpretation as a bounded signed correlation measure. More generally, the results indicate that the strength of cross-correlations may vary substantially across fluctuation magnitudes, highlighting the importance of amplitude-stratified analyses in complex systems.


The analysis of meteorological records further highlighted the ability of \(\rho_{\mathrm{SMFDCCA}}(n,q)\) to uncover heterogeneous coupling mechanisms that remain hidden to conventional correlation estimators. Such results demonstrate the capacity of the proposed observable to present amplitude-conditioned cross-correlation structures that remain hidden to conventional correlation estimators. More generally, the proposed observable extends the concept of detrended cross-correlation by introducing fluctuation amplitude as an explicit conditioning variable, thereby providing a complementary description of multiscale interactions that is inaccessible to conventional correlation observables.\\

\section*{Acknowledgments} 
This work received partial financial support from CNPq (Grant Numbers: $305096/2022-2$). Thiago B. Murari thanks National Council for Scientific and Technological Development --
CNPq (Process $444887/2024-5$) and FAPESB (Process $PIE0007/2022$). During the preparation of this work, the authors used Antigravity Plataform in order to evaluate the code in Python and test the scenarios. After using this tool, the authors reviewed and edited the content as needed and takes full responsibility for the content of the published article.

\section*{Data Availability}

The datasets, Python implementation, and example files supporting the findings of this study are publicly available in the Zenodo repository at \url{https://doi.org/10.5281/zenodo.21242106}. The repository contains the complete implementation of \(\rho_{\mathrm{SMFDCCA}}(n,q)\), together with the scripts required to reproduce the numerical analyses presented in this work.

\section*{End Matter}

\subsection*{Numerical Stability of the Regularization Parameter}

The computation of the Signed Multifractal Detrended Cross-Correlation Coefficient employs a small positive regularization parameter, $\varepsilon$, exclusively to prevent numerical singularities when evaluating negative fluctuation orders ($q<0$). The regularization is introduced only to ensure numerical robustness when local detrended fluctuation amplitudes approach zero and does not modify the mathematical definition of the proposed coefficient.

To evaluate the numerical sensitivity of the implementation, $\rho_{\mathrm{SMFDCCA}}(n,q)$ was recomputed for

\[
\varepsilon \in
\left\{
10^{-20},
10^{-18},
10^{-16},
10^{-14},
10^{-12},
10^{-10},
10^{-8},
10^{-6}
\right\},
\]

using all application datasets presented in this work, namely the Dow Jones Industrial Average versus NASDAQ Composite, mean temperature versus mean relative humidity in São Paulo, and maximum versus minimum temperature in São Paulo.

For each dataset, the complete $\rho_{\mathrm{SMFDCCA}}(n,q)$ surface was recomputed using the representative fluctuation orders

\[
q=-5,-2,0,2,5,
\]

and compared with the reference implementation using $\varepsilon=10^{-12}$. The Root Mean Square Error (RMSE), maximum absolute deviation, and mean deviation were evaluated for every tested value of $\varepsilon$.

No measurable numerical differences were observed over the entire investigated interval ($10^{-20}\leq\varepsilon\leq10^{-6}$). All recomputed coefficients were numerically indistinguishable from the reference implementation, with zero observed RMSE and zero maximum deviation across all datasets and fluctuation orders.

These results demonstrate that the proposed coefficient possesses a broad numerical stability plateau and is effectively insensitive to the precise choice of the regularization parameter over fourteen orders of magnitude. Consequently, $\varepsilon=10^{-12}$ was adopted as the default value throughout this work.

%


\begin{thebibliography}{21}%
\makeatletter
\providecommand \@ifxundefined [1]{%
 \@ifx{#1\undefined}
}%
\providecommand \@ifnum [1]{%
 \ifnum #1\expandafter \@firstoftwo
 \else \expandafter \@secondoftwo
 \fi
}%
\providecommand \@ifx [1]{%
 \ifx #1\expandafter \@firstoftwo
 \else \expandafter \@secondoftwo
 \fi
}%
\providecommand \natexlab [1]{#1}%
\providecommand \enquote  [1]{``#1''}%
\providecommand \bibnamefont  [1]{#1}%
\providecommand \bibfnamefont [1]{#1}%
\providecommand \citenamefont [1]{#1}%
\providecommand \href@noop [0]{\@secondoftwo}%
\providecommand \href [0]{\begingroup \@sanitize@url \@href}%
\providecommand \@href[1]{\@@startlink{#1}\@@href}%
\providecommand \@@href[1]{\endgroup#1\@@endlink}%
\providecommand \@sanitize@url [0]{\catcode `\\12\catcode `\$12\catcode `\&12\catcode `\#12\catcode `\^12\catcode `\_12\catcode `\%12\relax}%
\providecommand \@@startlink[1]{}%
\providecommand \@@endlink[0]{}%
\providecommand \url  [0]{\begingroup\@sanitize@url \@url }%
\providecommand \@url [1]{\endgroup\@href {#1}{\urlprefix }}%
\providecommand \urlprefix  [0]{URL }%
\providecommand \Eprint [0]{\href }%
\providecommand \doibase [0]{https://doi.org/}%
\providecommand \selectlanguage [0]{\@gobble}%
\providecommand \bibinfo  [0]{\@secondoftwo}%
\providecommand \bibfield  [0]{\@secondoftwo}%
\providecommand \translation [1]{[#1]}%
\providecommand \BibitemOpen [0]{}%
\providecommand \bibitemStop [0]{}%
\providecommand \bibitemNoStop [0]{.\EOS\space}%
\providecommand \EOS [0]{\spacefactor3000\relax}%
\providecommand \BibitemShut  [1]{\csname bibitem#1\endcsname}%
\let\auto@bib@innerbib\@empty
\bibitem [{\citenamefont {Di~Matteo}\ \emph {et~al.}(2025)\citenamefont {Di~Matteo}, \citenamefont {Moret}, \citenamefont {de~Barros~Pereira}, \citenamefont {Murari}, \citenamefont {da~Rocha~Filho},\ and\ \citenamefont {Mendes}}]{di2025epj}%
  \BibitemOpen
  \bibfield  {author} {\bibinfo {author} {\bibfnamefont {T.}~\bibnamefont {Di~Matteo}}, \bibinfo {author} {\bibfnamefont {M.~A.}\ \bibnamefont {Moret}}, \bibinfo {author} {\bibfnamefont {H.~B.}\ \bibnamefont {de~Barros~Pereira}}, \bibinfo {author} {\bibfnamefont {T.~B.}\ \bibnamefont {Murari}}, \bibinfo {author} {\bibfnamefont {T.~M.}\ \bibnamefont {da~Rocha~Filho}},\ and\ \bibinfo {author} {\bibfnamefont {J.~F.~F.}\ \bibnamefont {Mendes}},\ }\href@noop {} {\bibfield  {journal} {\bibinfo  {journal} {The European Physical Journal B}\ }\textbf {\bibinfo {volume} {98}},\ \bibinfo {pages} {171} (\bibinfo {year} {2025})}\BibitemShut {NoStop}%
\bibitem [{\citenamefont {Peng}\ \emph {et~al.}(1994)\citenamefont {Peng}, \citenamefont {Buldyrev}, \citenamefont {Havlin}, \citenamefont {Simons}, \citenamefont {Stanley},\ and\ \citenamefont {Goldberger}}]{peng1994mosaic}%
  \BibitemOpen
  \bibfield  {author} {\bibinfo {author} {\bibfnamefont {C.-K.}\ \bibnamefont {Peng}}, \bibinfo {author} {\bibfnamefont {S.~V.}\ \bibnamefont {Buldyrev}}, \bibinfo {author} {\bibfnamefont {S.}~\bibnamefont {Havlin}}, \bibinfo {author} {\bibfnamefont {M.}~\bibnamefont {Simons}}, \bibinfo {author} {\bibfnamefont {H.~E.}\ \bibnamefont {Stanley}},\ and\ \bibinfo {author} {\bibfnamefont {A.~L.}\ \bibnamefont {Goldberger}},\ }\href@noop {} {\bibfield  {journal} {\bibinfo  {journal} {Physical Review E}\ }\textbf {\bibinfo {volume} {49}},\ \bibinfo {pages} {1685} (\bibinfo {year} {1994})}\BibitemShut {NoStop}%
\bibitem [{\citenamefont {Podobnik}\ and\ \citenamefont {Stanley}(2008)}]{podobnik2008detrended}%
  \BibitemOpen
  \bibfield  {author} {\bibinfo {author} {\bibfnamefont {B.}~\bibnamefont {Podobnik}}\ and\ \bibinfo {author} {\bibfnamefont {H.~E.}\ \bibnamefont {Stanley}},\ }\href@noop {} {\bibfield  {journal} {\bibinfo  {journal} {Physical Review Letters}\ }\textbf {\bibinfo {volume} {100}},\ \bibinfo {pages} {084102} (\bibinfo {year} {2008})}\BibitemShut {NoStop}%
\bibitem [{\citenamefont {Kantelhardt}\ \emph {et~al.}(2002)\citenamefont {Kantelhardt}, \citenamefont {Zschiegner}, \citenamefont {Koscielny-Bunde}, \citenamefont {Havlin}, \citenamefont {Bunde},\ and\ \citenamefont {Stanley}}]{kantelhardt2002multifractal}%
  \BibitemOpen
  \bibfield  {author} {\bibinfo {author} {\bibfnamefont {J.~W.}\ \bibnamefont {Kantelhardt}}, \bibinfo {author} {\bibfnamefont {S.~A.}\ \bibnamefont {Zschiegner}}, \bibinfo {author} {\bibfnamefont {E.}~\bibnamefont {Koscielny-Bunde}}, \bibinfo {author} {\bibfnamefont {S.}~\bibnamefont {Havlin}}, \bibinfo {author} {\bibfnamefont {A.}~\bibnamefont {Bunde}},\ and\ \bibinfo {author} {\bibfnamefont {H.~E.}\ \bibnamefont {Stanley}},\ }\href@noop {} {\bibfield  {journal} {\bibinfo  {journal} {Physica A: Statistical Mechanics and its Applications}\ }\textbf {\bibinfo {volume} {316}},\ \bibinfo {pages} {87} (\bibinfo {year} {2002})}\BibitemShut {NoStop}%
\bibitem [{\citenamefont {Zhou}(2008)}]{zhou2008multifractal}%
  \BibitemOpen
  \bibfield  {author} {\bibinfo {author} {\bibfnamefont {W.-X.}\ \bibnamefont {Zhou}},\ }\href@noop {} {\bibfield  {journal} {\bibinfo  {journal} {Physical Review E}\ }\textbf {\bibinfo {volume} {77}},\ \bibinfo {pages} {066211} (\bibinfo {year} {2008})}\BibitemShut {NoStop}%
\bibitem [{\citenamefont {Zebende}(2011)}]{zebende2011dcca}%
  \BibitemOpen
  \bibfield  {author} {\bibinfo {author} {\bibfnamefont {G.~F.}\ \bibnamefont {Zebende}},\ }\href@noop {} {\bibfield  {journal} {\bibinfo  {journal} {Physica A: Statistical Mechanics and its Applications}\ }\textbf {\bibinfo {volume} {390}},\ \bibinfo {pages} {614} (\bibinfo {year} {2011})}\BibitemShut {NoStop}%
\bibitem [{\citenamefont {Nascimento~Filho}\ \emph {et~al.}(2018)\citenamefont {Nascimento~Filho}, \citenamefont {Pereira}, \citenamefont {Ferreira}, \citenamefont {Murari},\ and\ \citenamefont {Moret}}]{nascimento2018cross}%
  \BibitemOpen
  \bibfield  {author} {\bibinfo {author} {\bibfnamefont {A.}~\bibnamefont {Nascimento~Filho}}, \bibinfo {author} {\bibfnamefont {E.}~\bibnamefont {Pereira}}, \bibinfo {author} {\bibfnamefont {P.}~\bibnamefont {Ferreira}}, \bibinfo {author} {\bibfnamefont {T.}~\bibnamefont {Murari}},\ and\ \bibinfo {author} {\bibfnamefont {M.}~\bibnamefont {Moret}},\ }\href@noop {} {\bibfield  {journal} {\bibinfo  {journal} {Physica A: Statistical Mechanics and its Applications}\ }\textbf {\bibinfo {volume} {508}},\ \bibinfo {pages} {550} (\bibinfo {year} {2018})}\BibitemShut {NoStop}%
\bibitem [{\citenamefont {Ferreira}\ \emph {et~al.}(2020)\citenamefont {Ferreira}, \citenamefont {Kristoufek},\ and\ \citenamefont {Pereira}}]{ferreira2020dcca}%
  \BibitemOpen
  \bibfield  {author} {\bibinfo {author} {\bibfnamefont {P.}~\bibnamefont {Ferreira}}, \bibinfo {author} {\bibfnamefont {L.}~\bibnamefont {Kristoufek}},\ and\ \bibinfo {author} {\bibfnamefont {E.~J. d. A.~L.}\ \bibnamefont {Pereira}},\ }\href@noop {} {\bibfield  {journal} {\bibinfo  {journal} {Physica A: Statistical Mechanics and its Applications}\ }\textbf {\bibinfo {volume} {545}},\ \bibinfo {pages} {123803} (\bibinfo {year} {2020})}\BibitemShut {NoStop}%
\bibitem [{\citenamefont {Azevedo}\ \emph {et~al.}(2021)\citenamefont {Azevedo}, \citenamefont {Sampaio}, \citenamefont {Filho}, \citenamefont {Moret},\ and\ \citenamefont {Murari}}]{azevedo2021sustainable}%
  \BibitemOpen
  \bibfield  {author} {\bibinfo {author} {\bibfnamefont {G.~A.}\ \bibnamefont {Azevedo}}, \bibinfo {author} {\bibfnamefont {R.~R.}\ \bibnamefont {Sampaio}}, \bibinfo {author} {\bibfnamefont {A.~S.~N.}\ \bibnamefont {Filho}}, \bibinfo {author} {\bibfnamefont {M.~A.}\ \bibnamefont {Moret}},\ and\ \bibinfo {author} {\bibfnamefont {T.~B.}\ \bibnamefont {Murari}},\ }\href@noop {} {\bibfield  {journal} {\bibinfo  {journal} {Scientific reports}\ }\textbf {\bibinfo {volume} {11}},\ \bibinfo {pages} {791} (\bibinfo {year} {2021})}\BibitemShut {NoStop}%
\bibitem [{\citenamefont {Oliveira}\ \emph {et~al.}(2023)\citenamefont {Oliveira}, \citenamefont {Murari}, \citenamefont {Nascimento~Filho}, \citenamefont {Saba}, \citenamefont {Moret},\ and\ \citenamefont {Cardoso}}]{oliveira2023paradox}%
  \BibitemOpen
  \bibfield  {author} {\bibinfo {author} {\bibfnamefont {J.~B.}\ \bibnamefont {Oliveira}}, \bibinfo {author} {\bibfnamefont {T.~B.}\ \bibnamefont {Murari}}, \bibinfo {author} {\bibfnamefont {A.~S.}\ \bibnamefont {Nascimento~Filho}}, \bibinfo {author} {\bibfnamefont {H.}~\bibnamefont {Saba}}, \bibinfo {author} {\bibfnamefont {M.~A.}\ \bibnamefont {Moret}},\ and\ \bibinfo {author} {\bibfnamefont {C.~A.~L.}\ \bibnamefont {Cardoso}},\ }\href@noop {} {\bibfield  {journal} {\bibinfo  {journal} {Science of The Total Environment}\ }\textbf {\bibinfo {volume} {860}},\ \bibinfo {pages} {160491} (\bibinfo {year} {2023})}\BibitemShut {NoStop}%
\bibitem [{\citenamefont {Figueredo}\ \emph {et~al.}(2023)\citenamefont {Figueredo}, \citenamefont {Monteiro}, \citenamefont {do~Nascimento~Silva}, \citenamefont {de~Ara{\'u}jo~Fontoura}, \citenamefont {da~Silva},\ and\ \citenamefont {Alves}}]{figueredo2023analysis}%
  \BibitemOpen
  \bibfield  {author} {\bibinfo {author} {\bibfnamefont {M.~B.}\ \bibnamefont {Figueredo}}, \bibinfo {author} {\bibfnamefont {R.~L.~S.}\ \bibnamefont {Monteiro}}, \bibinfo {author} {\bibfnamefont {A.}~\bibnamefont {do~Nascimento~Silva}}, \bibinfo {author} {\bibfnamefont {J.~R.}\ \bibnamefont {de~Ara{\'u}jo~Fontoura}}, \bibinfo {author} {\bibfnamefont {A.~R.}\ \bibnamefont {da~Silva}},\ and\ \bibinfo {author} {\bibfnamefont {C.~A.~P.}\ \bibnamefont {Alves}},\ }\href@noop {} {\bibfield  {journal} {\bibinfo  {journal} {Scientific Reports}\ }\textbf {\bibinfo {volume} {13}},\ \bibinfo {pages} {7512} (\bibinfo {year} {2023})}\BibitemShut {NoStop}%
\bibitem [{\citenamefont {Kwapie{\'n}}\ \emph {et~al.}(2015)\citenamefont {Kwapie{\'n}}, \citenamefont {O{\'s}wi{\k{e}}cimka},\ and\ \citenamefont {Dro{\.z}d{\.z}}}]{kwapien2015detrended}%
  \BibitemOpen
  \bibfield  {author} {\bibinfo {author} {\bibfnamefont {J.}~\bibnamefont {Kwapie{\'n}}}, \bibinfo {author} {\bibfnamefont {P.}~\bibnamefont {O{\'s}wi{\k{e}}cimka}},\ and\ \bibinfo {author} {\bibfnamefont {S.}~\bibnamefont {Dro{\.z}d{\.z}}},\ }\href@noop {} {\bibfield  {journal} {\bibinfo  {journal} {Physical Review E}\ }\textbf {\bibinfo {volume} {92}},\ \bibinfo {pages} {052815} (\bibinfo {year} {2015})}\BibitemShut {NoStop}%
\bibitem [{\citenamefont {da~Silva~Filho}\ \emph {et~al.}(2026)\citenamefont {da~Silva~Filho}, \citenamefont {de~Castro},\ and\ \citenamefont {Guedes}}]{silvafilho2026multifractal}%
  \BibitemOpen
  \bibfield  {author} {\bibinfo {author} {\bibfnamefont {A.~M.}\ \bibnamefont {da~Silva~Filho}}, \bibinfo {author} {\bibfnamefont {A.~P.~N.}\ \bibnamefont {de~Castro}},\ and\ \bibinfo {author} {\bibfnamefont {E.~F.}\ \bibnamefont {Guedes}},\ }\href@noop {} {\bibfield  {journal} {\bibinfo  {journal} {Physica A: Statistical Mechanics and its Applications}\ }\textbf {\bibinfo {volume} {689}},\ \bibinfo {pages} {131424} (\bibinfo {year} {2026})}\BibitemShut {NoStop}%
\bibitem [{\citenamefont {Sierra-Porta}(2025)}]{sierraporta2025multifractal}%
  \BibitemOpen
  \bibfield  {author} {\bibinfo {author} {\bibfnamefont {D.}~\bibnamefont {Sierra-Porta}},\ }\href@noop {} {\bibfield  {journal} {\bibinfo  {journal} {Journal of Atmospheric and Solar-Terrestrial Physics}\ }\textbf {\bibinfo {volume} {266}},\ \bibinfo {pages} {106407} (\bibinfo {year} {2025})}\BibitemShut {NoStop}%
\bibitem [{\citenamefont {W{\k{a}}torek}\ \emph {et~al.}(2021)\citenamefont {W{\k{a}}torek}, \citenamefont {Dro{\.z}d{\.z}}, \citenamefont {Kwapie{\'n}}, \citenamefont {Minati}, \citenamefont {O{\'s}wi{\k{e}}cimka},\ and\ \citenamefont {Stanuszek}}]{watorek2021}%
  \BibitemOpen
  \bibfield  {author} {\bibinfo {author} {\bibfnamefont {M.}~\bibnamefont {W{\k{a}}torek}}, \bibinfo {author} {\bibfnamefont {S.}~\bibnamefont {Dro{\.z}d{\.z}}}, \bibinfo {author} {\bibfnamefont {J.}~\bibnamefont {Kwapie{\'n}}}, \bibinfo {author} {\bibfnamefont {L.}~\bibnamefont {Minati}}, \bibinfo {author} {\bibfnamefont {P.}~\bibnamefont {O{\'s}wi{\k{e}}cimka}},\ and\ \bibinfo {author} {\bibfnamefont {M.}~\bibnamefont {Stanuszek}},\ }\href@noop {} {\bibfield  {journal} {\bibinfo  {journal} {Physics Reports}\ }\textbf {\bibinfo {volume} {901}},\ \bibinfo {pages} {1} (\bibinfo {year} {2021})}\BibitemShut {NoStop}%
\bibitem [{\citenamefont {Jiang}\ \emph {et~al.}(2019)\citenamefont {Jiang}, \citenamefont {Xie}, \citenamefont {Zhou},\ and\ \citenamefont {Sornette}}]{Jiang2019}%
  \BibitemOpen
  \bibfield  {author} {\bibinfo {author} {\bibfnamefont {Z.-Q.}\ \bibnamefont {Jiang}}, \bibinfo {author} {\bibfnamefont {W.-J.}\ \bibnamefont {Xie}}, \bibinfo {author} {\bibfnamefont {W.-X.}\ \bibnamefont {Zhou}},\ and\ \bibinfo {author} {\bibfnamefont {D.}~\bibnamefont {Sornette}},\ }\href@noop {} {\bibfield  {journal} {\bibinfo  {journal} {Reports on Progress in Physics}\ }\textbf {\bibinfo {volume} {82}},\ \bibinfo {pages} {125901} (\bibinfo {year} {2019})}\BibitemShut {NoStop}%
\bibitem [{\citenamefont {Di~Matteo}(2007)}]{DiMatteo2007}%
  \BibitemOpen
  \bibfield  {author} {\bibinfo {author} {\bibfnamefont {T.}~\bibnamefont {Di~Matteo}},\ }\href@noop {} {\bibfield  {journal} {\bibinfo  {journal} {Quantitative Finance}\ }\textbf {\bibinfo {volume} {7}},\ \bibinfo {pages} {21} (\bibinfo {year} {2007})}\BibitemShut {NoStop}%
\bibitem [{\citenamefont {Barunik}\ \emph {et~al.}(2012)\citenamefont {Barunik}, \citenamefont {Aste}, \citenamefont {Di~Matteo},\ and\ \citenamefont {Liu}}]{barunik2012understanding}%
  \BibitemOpen
  \bibfield  {author} {\bibinfo {author} {\bibfnamefont {J.}~\bibnamefont {Barunik}}, \bibinfo {author} {\bibfnamefont {T.}~\bibnamefont {Aste}}, \bibinfo {author} {\bibfnamefont {T.}~\bibnamefont {Di~Matteo}},\ and\ \bibinfo {author} {\bibfnamefont {R.}~\bibnamefont {Liu}},\ }\href@noop {} {\bibfield  {journal} {\bibinfo  {journal} {Physica A: Statistical Mechanics and its Applications}\ }\textbf {\bibinfo {volume} {391}},\ \bibinfo {pages} {4234} (\bibinfo {year} {2012})}\BibitemShut {NoStop}%
\bibitem [{\citenamefont {Mantegna}(1999)}]{Mantegna1999}%
  \BibitemOpen
  \bibfield  {author} {\bibinfo {author} {\bibfnamefont {R.~N.}\ \bibnamefont {Mantegna}},\ }\href@noop {} {\bibfield  {journal} {\bibinfo  {journal} {The European Physical Journal B-Condensed Matter and Complex Systems}\ }\textbf {\bibinfo {volume} {11}},\ \bibinfo {pages} {193} (\bibinfo {year} {1999})}\BibitemShut {NoStop}%
\bibitem [{\citenamefont {Epps}(1979)}]{epps1979comovements}%
  \BibitemOpen
  \bibfield  {author} {\bibinfo {author} {\bibfnamefont {T.~W.}\ \bibnamefont {Epps}},\ }\href@noop {} {\bibfield  {journal} {\bibinfo  {journal} {Journal of the American Statistical Association}\ }\textbf {\bibinfo {volume} {74}},\ \bibinfo {pages} {291} (\bibinfo {year} {1979})}\BibitemShut {NoStop}%
\bibitem [{\citenamefont {Dro{\.z}d{\.z}}\ \emph {et~al.}(2010)\citenamefont {Dro{\.z}d{\.z}}, \citenamefont {Kwapie{\'n}}, \citenamefont {O{\'s}wi{\k{e}}cimka},\ and\ \citenamefont {Rak}}]{drozdz2010foreign}%
  \BibitemOpen
  \bibfield  {author} {\bibinfo {author} {\bibfnamefont {S.}~\bibnamefont {Dro{\.z}d{\.z}}}, \bibinfo {author} {\bibfnamefont {J.}~\bibnamefont {Kwapie{\'n}}}, \bibinfo {author} {\bibfnamefont {P.}~\bibnamefont {O{\'s}wi{\k{e}}cimka}},\ and\ \bibinfo {author} {\bibfnamefont {R.}~\bibnamefont {Rak}},\ }\href@noop {} {\bibfield  {journal} {\bibinfo  {journal} {New Journal of Physics}\ }\textbf {\bibinfo {volume} {12}},\ \bibinfo {pages} {105003} (\bibinfo {year} {2010})}\BibitemShut {NoStop}%
\bibitem{INMET_BDMEP}
Instituto Nacional de Meteorologia (INMET),
\textit{Banco de Dados Meteorológicos para Ensino e Pesquisa (BDMEP)},
\url{https://bdmep.inmet.gov.br/} (accessed June 1, 2026).
\end{thebibliography}
\end{document}